\documentclass[12pt,preprint]{aastex}

\usepackage{epsfig, natbib}

\newcommand{\siml}{\lower4pt \hbox{$\buildrel < \over \sim$}}
\newcommand{\simg}{\lower4pt \hbox{$\buildrel > \over \sim$}}

\newcommand{\swift}{{\it Swift}}

\newcommand{\agile}{AGILE}
\newcommand{\glast}{GLAST}

\def\beq{\begin{equation}}
\def\enq{\end{equation}}
\def\bea{\begin{array}}
\def\ena{\end{array}}

\begin{document}

\title{GLAST Prospects for Swift-Era Afterglows}

\author{ L.~J. Gou\altaffilmark{1} and P. M\'{e}sz\'{a}ros\altaffilmark{1,2}}

\altaffiltext{1}{Department of Astronomy \& Astrophysics, 525 Davey
Laboratory, Pennsylvania State University, University Park, PA
16802}

\altaffiltext{2}{Department of Physics, Pennsylvania State
University, University Park, PA 16802}

\begin{abstract}

We calculate the GeV spectra of GRB afterglows produced by inverse Compton
scattering of the sub-MeV emission of these objects. We improve on earlier
treatments by using refined afterglow parameters and new model developments
motivated by recent Swift observations. We present time-dependent GeV
spectra for standard, constant parameter models, as well as for models with
energy injection and with time-varying parameters, for a range of burst 
parameters. We evaluate the limiting redshift to which such afterglows
can be detected by the GLAST LAT, as well as AGILE.

\end{abstract}

\section{Introduction}
\label{sec:intro}

GRB afterglow observations from X-rays to radio are generally well
described by an external shock model \citep[e.g.][for a recent
review]{mes+06}. However, model fits to the data still leave
uncertainties in some of the parameters of the basic external shock
model, and the bolometric luminosity depends on the poorly known GeV
range spectrum. The Gamma Ray Large Area Space Telescope GLAST
\citep{mce+06} is expected to be launched at the end of 2007. The
Large Area Telescope (LAT) covers the energy range from 20 MeV to 300
GeV, while the Gamma Ray Burst Monitor (GBM) covers the range from 8
keV to 25 MeV. The effective area of the LAT is about 7 times larger
than of the previous EGRET experiment at GeV energies.
\agile\footnote{http://agile.rm.iasf.cnr.it/index.html} was
successfully launched on April 23, 2007, with an energy range of 30
MeV to 50 GeV \citep{tbab+06}.  It is hoped that the higher photon
statistics at GeV energies of GLAST and AGILE will lead tighter
constraints on the burst parameters, and an improved understanding of
the GeV afterglow spectra.

GLAST may also be able to test recent developments in the
understanding of GRB afterglows, motivated by observations with the
Swift satellite \citep[e.g.][]{nkgp+06,zfdk+06,fm+06,zb+07}.
One such development is the presence of a shallow decay phase of the
X-ray afterglow, which may be due to refreshed shocks or late energy
injection \citep[e.g.][]{zfdk+06}, or alternatively, it may be due to
a change in time of the shock parameters \citep[e.g.][]{ityn+06}. We
investigate here the effects of such new features on the expected GeV
spectrum.

Another question of great interest is how far can GLAST detect the MeV
to GeV emission from such bursts, both in the basic model and the case
where new features such as the above are present. This requires a
detailed calculation of the GeV spectrum as a function of time, with
allowance for the changes in the dynamics implied by the injection,
time variability, etc. The most conservative and widely considered
mechanism for producing photons in this range is inverse Compton (IC)
scattering
\citep{pm+98,t+98,wl+98,cd+99,pk+00,se+01,zm+01,wdl+01b,wlm+06,wf+07,fpnw+07,gp+07}.
Another potential mechanism is hadronic cascades following proton
acceleration \citep{bd+98,zm+01,fmpt+04,gz+07}.  This mechanism is
less well constrained, depending on the efficiency of proton
acceleration; it may be important in the prompt phase, but its
parameter regime is small and generally outside of the afterglow
parameter fit range \citep{zm+01}, so it is not considered here.  The
maximum distance to which GLAST could detected GRB afterglow was
discussed by \citet{zm+01} for the basic standard model, using a
simplified analytical broken power-law approximation to the IC
spectrum. This resulted in an IC-to-synchrotron peak-flux ratio which
is overestimated by a factor $\sim 10$, compared to a more accurate
calculation, as we discuss in this paper. The usefulness of this ratio
is that it allows simple predictions for the detectability in the GeV
range based on the lower energy measurements. Here we discuss the
validity of the simple analytical approximations, compared to more
accurate numerical calculations of afterglow spectra at GeV energies.

In $\S$ \ref{sec:model} we describe the afterglow synchrotron-inverse
Compton model used for the numerical computations, a comparison
between numerical and analytical approximations being given in the
Appendix.  In $\S$ \ref{sec:comp_detection} we then present both
numerical IC spectra and their appropriate analytical approximations,
for the basic GRB afterglow model and for the extended models
including new Swift-motivated elements such as energy injection or
time-varying parameters, and evaluate their detectability as a
function of redshift for the GLAST LAT instrument and for \agile.

\section{Afterglow Synchrotron-inverse Compton Spectra at GeV Energies}
\label{sec:model}

The afterglow of a GRB, due to the external shock as it slows down in
the external medium, produces synchrotron radiation in the X-ray to
MeV range, which is then inverse-Compton upscattered into the GeV-TeV
range \citep{mrp+94,bd+98}. More specific calculation of the IC GeV
range are, e.g. those of \citet{se+01}, \citet{zm+01}, \citet{pw+04},
etc.  We describe the afterglow models, as usual, by the total
isotropic energy $E_{52,iso}=E/10^{52}{\rm erg}$ (for the case of
energy injection see below), the external density $n$, a jet opening
half-angle $\theta$, electron equipartition parameter $\epsilon_e$,
magnetic equipartition parameter $\epsilon_B$ and electron energy
index $p$. The other parameter of relevance in synchrotron-IC models
is the scattering Y-parameter which is defined as the luminosity ratio
of IC to synchrotron, usually given by
\begin{equation}
Y=(-1+\sqrt{4\epsilon_e/\epsilon_B+1})/2 ~.
\label{eq:y-param1}
\end{equation}
The initial Lorentz factor $\Gamma_0$ of the burst is not needed as a
parameter, since in the asymptotic blast wave regime the Lorentz
factor follows from the scaling law,
\begin{equation}
\Gamma (t) = (17E/1024 \pi n m_p c^5 t^3)^{1/8}
\label{eq:Gamma}
\end{equation}
Our numerical calculations of the spectra and fluence curves use the
basic synchrotron-IC equations given in \citet{gfm+06}, extending now
to the GeV range.  We also consider in this range the spectral effects
of the photon-photon opacity effects, which impose cutoffs depending
on the spectrum and density of lower energy photons \citep{bh+97,
ls+01}.  

We calculate the synchrotron-IC spectra of three different types of
GRB afterglow models, and evaluate their detectability with the GLAST
Large Area Telescope (LAT) and with AGILE. These calculations improve
on previous calculations, e.g. \citep{zm+01}, in several
respects. First, in the ``standard'' afterglow model such as used in
\citep{zm+01}, we use the exact IC spectrum instead of the broken
power-law approximation, and the peak flux ratio is taken as F1
instead of F2 (see Appendix).  Second, we include the complete
spectral regimes, not only the commonly used $\nu_a < (\nu_m, \nu_c)$
regimes where $\nu_a$, $\nu_m$, and $\nu_c$ are the synchrotron
self-absorption, injection, and cooling frequencies, respectively
\citep{gfm+06}. This assures that the GRB afterglows evolve through
the correct regimes at all times.  Third, we consider Swift-motivated
additions to the standard afterglow model, such as a continued energy
injection model, and a varying afterglow parameter model (motivated by
the presence of a shallow decay phase, and a high apparent radiative
efficiency, see e.g. \citeauthor{mes+06}\,\citeyear{mes+06} for a
review).

The details of the three models calculated are as follows.  (A) A
standard afterglow model, with all parameters remaining constant
during its evolution (for a detailed description of this model, see
\citeauthor{gfm+06}\,\citeyear{gfm+06}).  (B) A continuous energy
injection model, which is a widely considered model to explain the
shallow decay phase commonly seen in \swift\, light curves. For this,
we assume that the total kinetic energy increases over time with a
power-law index, $E\propto t^{1-q}$, before the break time $t=10^4$
seconds and the break time here is defined as the one when the shallow
decay phase ends. Fits to Swift observations indicate a value $q\sim
0.5$~\citep{zfdk+06}. (C) An evolving parameter model, which is an
alternative model for explaining the shallow decay phase, based on the
assumption that the electron equipartition parameter $\epsilon_e$
increases with the time as $\propto t^{\alpha}$ \citep{ityn+06} before
the break time, as for the energy injection model. For all three
models, we assume that they have the same parameters at late times,
i.e.  after the break time.  Since the flux has to be integrated over
the observation time, we set the observation time to be one half of
the final time of observation since the trigger (e.g. if the
observational data stop at $t=10^5$ seconds, the integration time is
from $t=5\times 10^4$ seconds to $t=10^5$ seconds). This is consistent
with the \glast\, observation characteristics, as well as those of
\agile, in the point-source observing mode, where due to earth
occultation, only about $50\%$ of the orbit time is used for the burst
observation.

To determine the limiting redshift to which a burst can be detected,
we calculate the instrumental fluence threshold as in \citet{zm+01},
using the instrument characteristics of the GLAST LAT and \agile.  For
a flux sensitivity $\Phi_m~{\rm ph~cm^{-2}~s^{-1}}$ over an exposure
time T and a point source observed over an effective observation time
$t_{eff}$ (in seconds) in an energy band centered around a photon
energy $E$, the fluence threshold is estimated as $F_{thr}\sim [\Phi_o
(T/t_{eff})^{1/2}] E t_{eff} $ where $[\Phi_o (T/t_{eff})^{1/2}]$ is
the sensitivity for the effective observation time $t_{eff}$ because
the sensitivity scales as $\sqrt{t_{eff}}$ for the longer observations
where the sensitivity is limited by the background.  Due to the
occultation by the earth, the effective observation time $t_{eff}$ is
normally $\le 50\%$ of the total orbit time, $t_{obs}$, for both GLAST
and \agile ~ (or equivalent to the observation time after the burst),
namely, $t_{eff}=\eta t_{obs}$ where the observing efficiency $\eta$
is taken to be $\eta=0.5$. Hereafter, unless otherwise stated, the
small ``$t$'' without the subscript refers to the observation time
$t_{obs}$ for simplicity. For GLAST we use the fluence threshold
listed in the updated instrument performance documents\footnote{
http://www-glast.slac.stanford.edu/software/IS/glast\_lat\_performance.htm}.
Considering the integral sensitivity above 100 MeV for GLAST LAT to be
$\sim 4\times 10^{-9} {\rm ph~cm^{-2}~s^{-1}}$ for an effective
observation time of one year in the sky-survey mode, the fluence
threshold is $1.0\times 10^{-8}t^{1/2}~{\rm erg~ cm^{-2}}$ for the
long-time observation in the sky-survey mode. For GRB afterglows, in
most cases GLAST will perform a pointed observation rather than the
survey mode observation. In this mode, \glast\, will keep the GRB
position always at the center of the LAT field of view for as long as
possible, and this will improve the sensitivity by a factor of 3-5
(depending on where the GRB lies with respect to the orbital plane; an
object which lies at the orbit pole will not be occulted by the Earth
and can be continuously observed; J. McEnery 2007, private
communication). Therefore, taking the improvement factor of 3, the
fluence threshold for a GRB observation is $3.4\times 10^{-9}
t^{1/2}~{\rm erg~ cm^{-2}}$ for the long-time observation. The
short-time fluence threshold can be defined by the criterion that at
least 5 photons are collected and it depends on the effective area of
the instrument which is around 6000 $\rm cm^2$ at 400 MeV for \glast\,
LAT, so it is $5.3\times 10^{-7}~{\rm erg~cm^{-2}}$ (the transition
time when the short-time sensitivity and long-time sensitivity meet is
$2.4\times 10^4$\,seconds). Compared to the previous estimate of
\citet{zm+01} for the GLAST sensitivity, the short-time observation
sensitivity here is roughly the same, but the long-time sensitivity
has changed from $1.2\times 10^{-9}t^{1/2}$ to $3.4\times
10^{-9}t^{1/2}~{\rm erg~ cm^{-2}}$, increased by a factor of 3.

The energy range of \agile~ is somewhat lower than that of GLAST LAT.
Its flux sensitivity above 100 MeV is $\sim 3 \times 10^{-7}~{\rm
ph~cm^{-2}~s^{-1}}$ for a point-source observation over an effective
period of $10^6$ seconds \citep{tbab+06}. Thus, taking the observing
efficiency $\eta=0.5$, the fluence threshold can be estimated as
$3.0\times 10^{-7} (T/t_{eff})^{1/2} E t_{eff} \approx 1.3 \times
10^{-7}t^{1/2}~{\rm ergs~cm^{-2}}$ at an average energy of 400 MeV,
where $T=10^6$ seconds is the exposure time corresponding to the given
sensitivity. The fluence threshold is for the long-time
observation. The shorter-time fluence threshold can be obtained
similarly as above, $\sim (5/550)\times 400$ MeV $\sim 5.8\times
10^{-6}~{\rm ergs~cm^{-2}}$ where we have taken the effective area for
\agile\, to be 550 $\rm cm^2$.  In summary, for \agile, the fluence
threshold for the long-time observation (i.e., $t> 1870$ seconds) is
$1.3 \times 10^{-7}t^{1/2}~{\rm ergs~cm^{-2}}$ and $5.8\times 10^{-6}~
{\rm ergs~cm^{-2}}$ for the short-time observation, the transition
time being $\simeq 1870$ seconds.

\section{Detectability of GRB Afterglows with GLAST and AGILE}
\label{sec:comp_detection}

The initial nominal set of parameters for the standard model (A) used here 
are the same as for the standard model of \citet{zm+01}: $p=2.2,
\epsilon_e=0.5, \epsilon_B=0.01, E_{52,iso}=1,n=1~{\rm cm^{-3}}$. An
additional feature is that we also assume a jet opening half-angle
$\theta=0.14$, which does not affect the flux at early times.  The
other parameters are as for model (A). For the injection model (B),
the kinetic energy is assumed to increase following the relation
$E\propto t^{1-q}=t^{0.5}$ where $q$ is taken to be 0.5 based on the
\swift\, observation fits \citep{zfdk+06}. For the model (C) with
evolving parameters, we assumed $\epsilon_e$ to follow the relation
$\epsilon_e\propto t^{0.5}$ \citep{ityn+06}, other parameters being
the same as for model (A). In the alternative models (B) and (C),
either the kinetic energy or the electron equipartition parameter
starts out with a smaller value as for (A), but at late times end up
with the same values as the standard model (A). The transition time at
which the energy injection or the $\epsilon_e$ evolution stops is set
at $t=10^4$ seconds.

\begin{figure}
\plotone{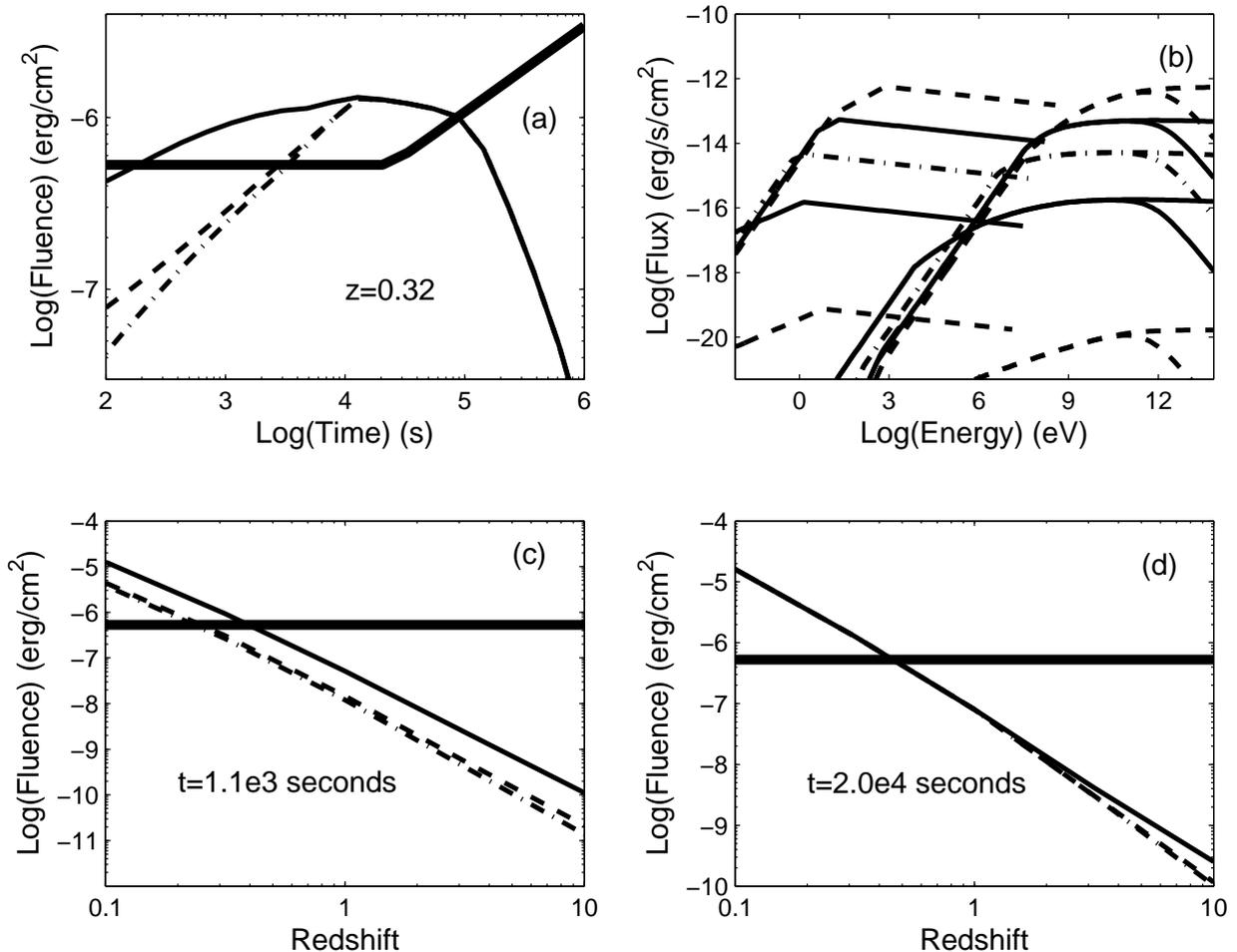}
\caption{{\it Panel (a)} The partial fluence curves (defined as flux
integrated between $0.5t$ and $t$) as a function of the observation
time $t$ since the trigger, for the three GRB afterglow models in the
GLAST LAT energy band, at a redshift $z=0.32$.  Thin solid curves:
standard model (A) (constant energy); dashed curves: ``energy
injection'' model (B); dot-dashed curves: ``evolving parameter'' model
(C).  The sensitivity of the GLAST LAT is shown by the thick broken
solid curve.  The parameters for the standard model are
$E_{52,iso}=1,~\epsilon_e=0.5$, $\epsilon_{B}=0.01$, $p=2.2,
\theta=0.14$ and the break time when the shallow decay phase ends is
at $t=10^4~{\rm seconds}$.  Before the break time, $\epsilon_e \propto
t^{0.5}$ in the evolving parameter model, and the kinetic energy
$E\propto t^{0.5}$ in the energy injection model. After the break
time, those parameters will be the same as the ones in the standard
model.  {\it Panel (b)} The synchrotron and IC spectrum for the
standard model at different time epochs: $10^2, 10^3, 10^4, 10^5$, and
$10^6$ seconds, respectively. Above photon energies $\sim 10^{12}$ eV,
the upper spectral curve represents the flux without $\gamma$-$\gamma$
absorption, and the lower curve is with inclusion of this absorption.
{\it Panel (c)} The redshift dependence of the partial fluence for the
three models above, evaluated at the observation time $t\sim 1.1\times
10^3$ seconds.  The thick solid line is the GLAST sensitivity for an
integration time 550 seconds (the observing efficiency 0.5 has been
taken).  The intersection of the partial fluence and the sensitivity
curve gives the limiting redshift, to which the bursts can be detected
by GLAST for this integration time, which is is $z\simeq 0.4$ for the
standard model, and $z \simeq 0.22$ for the other two models. {\it
Panel (d)} The redshift dependence of the partial fluence for the
three models above, evaluated at $t\sim 2.0\times 10^4$ seconds. The
thick solid line is the GLAST sensitivity for an integration time
$10^4$ seconds.  The intersection gives a limiting redshift for
detection of $z\simeq 0.45$ for all three models.}
\label{fig:zhang_detection}
\end{figure}

Fig \ref{fig:zhang_detection} shows the results for the three models
above, using the nominal set of parameters. Panel (a) shows the
partial fluence, defined here as the energy flux integrated over the
time intervals [$t-\Delta t$, $t$], as a function of $t$, where the
$t=t_{obs}$ is the observation time counted after the trigger,
adopting a nominal integration time $\Delta t=0.5 t$ throughout.  The
partial fluence curves shown correspond to the three different GRB
afterglow models, (A) the standard model, which is the same as the
type II GRB model in \citet{zm+01}; (B) the energy injection model;
(C) the evolving parameter model. As can be seen in panel (a), for a
burst at a low redshift $z=0.32$ the GeV emission from all three
models can be detected by GLAST up to a time $t\sim 1.5 \times 10^{5}$
seconds (the thick solid line indicates the \glast\, LAT
sensitivity). Note that the GeV emission from the standard model is
higher than that from the two other models.  This is because all the
models end up with the same energy and same parameters at late times,
which means the injection starts with lower energy and the evolving
parameter begins with the lower $\epsilon_e$ at the beginning.  Panel
(b) shows the synchrotron and IC spectra of the standard model (A) at
times $t=10^2, 10^3, 10^4, 10^5,$ and $10^6$ seconds. Hereafter,
unless otherwise stated, we always calculate the spectra at these time
epoches.  The fluxes around TeV ($10^{12}$ eV) show the effects of
inclusion of the photon-photon absorption within the sources. The
upper curves are the flux without the $\gamma$-$\gamma$ absorption,
and the lower curves are the flux after internal absorption. For this
we have used the optical depth to internal $\gamma$-$\gamma$
interactions of Equation (20) in \citet{zm+01}.  For the relatively
low compactness parameters of the external afterglow shock discussed
here, the $\gamma\gamma$ cutoff becomes important above $\sim$ TeV
energies, which is more of interest for ground-based air Cherenkov
telescope observations than for space detectors.  The lower panels (c)
and (d) show the redshift dependence of the GeV emission for all three
models, at $t\sim 1.1\times 10^3$ and $t\sim 2\times 10^4$ seconds.
At $t\sim 1.1\times 10^3$ seconds, the limiting redshift is $z\sim
0.4$ for the standard model and $z\sim 0.22$ for the other two
models. At $t\sim 2\times 10^4$, the limiting redshift is around
$z\sim 0.45$ for all the models.

Note that while the usual fluence is defined as flux integrated 
over the observation time since the trigger, which always increase with 
time, the partial fluences shown in panel (a) first increase and eventually
decrease. This is because the afterglow flux decreases with time $t$,
and for the partial fluences the integration time starts at $0.5 t$ 
and ends at $t$. This is done to check the limiting redshift to which 
afterglows can be detected for typical observations at different epochs 
$t$ with some uniform criterion for the integration time. The snapshot
at the time $2\times 10^4$ seconds lies where the partial fluence
is roughly flat in time, during which period the limiting redshift 
reaches its maximum (although the partial fluence within the flat 
phase varies by a factor $\le 2$, the limiting redshift changes only 
slightly). Other snapshot epochs were chosen around one decade 
earlier or later than the typical maximum redshift epoch.

\begin{figure}
\plotone{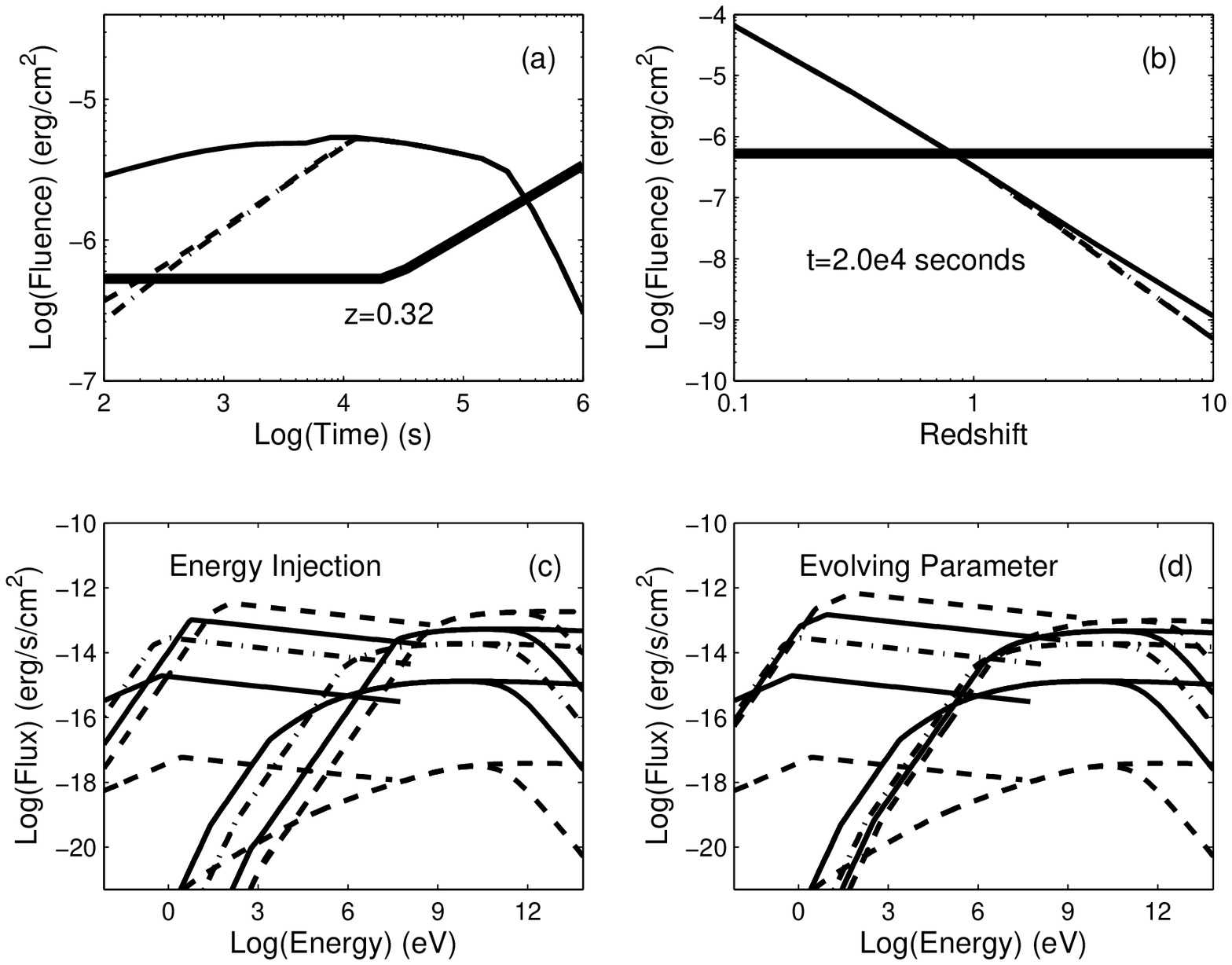}
\caption{{\it Panel (a)} The partial fluence curves of the same three
GRB afterglow models over the \glast\, energy band, for a different
set of parameters generally assumed to be 'typical', namely kinetic
energy $E_{52,iso}=10$,~ $\epsilon_e=0.2$ and other parameters the
same as in Fig (\ref{fig:zhang_detection}). {\it Panel (b)} The
redshift dependence of the partial fluence at $t=2\times 10^4$ s for
the same models, giving a limiting redshift $z\simeq 0.8$. {\it Panel
(c)} The synchrotron-IC spectra for the energy injection case. {\it
Panel (d)} The synchrotron-IC spectra for the evolving parameter
case.}
\label{fig:normal_energy}
\end{figure}

In Fig \ref{fig:normal_energy} we consider an alternative set of
parameters. The motivation for this is that the parameters of the
standard model shown in Figure \ref{fig:zhang_detection},
$E_{52,iso}=1$, and $\epsilon_e=0.5$, differ somewhat from the
'statistical average' values quoted for low-redshift GRBs,
$E_{52,jet}\sim 0.1$ and $\epsilon_e\sim 0.1$, e.g. \citet{pk+01}.  In
order to check the sensitivity of the detectability of GRBs to
variations in these parameters, we perform the same calculation suing
the values $p=2.2, \epsilon_e=0.2, \epsilon_B=0.01, E_{52,iso}=10,
n=1~{\rm cm^{-3}}$, and $\theta=0.14$, the results being shown in Fig
(\ref{fig:normal_energy}). For these 'average' parameters, the
limiting redshift is $z\simeq 0.8$ for all three GRB models at $t\sim
2.0 \times 10^4$ seconds.

\begin{figure}
\plotone{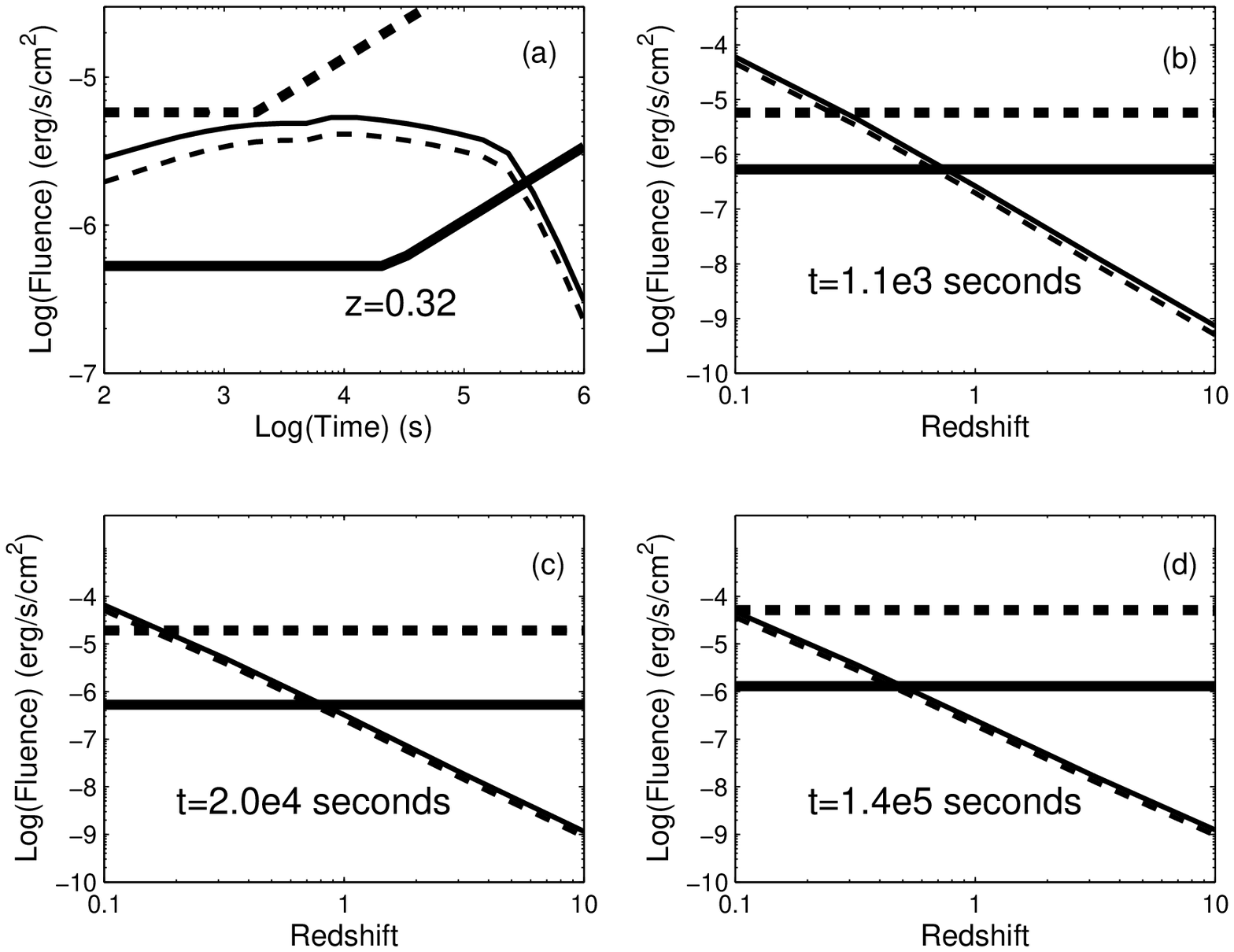}
\caption{Comparison of \glast\, and \agile\, detectability, for the
standard model with the parameters of Fig. \ref{fig:normal_energy}.  
{\it Panel (a)} The partial fluence curves for the standard model over the
energy bands of \agile\, (thin dashed line) and \glast\, (thin solid
line), compared to the respective instrument sensitivities (thick
dashed line for \agile\, and thick solid line for \glast). The
parameters are the same as in Fig \ref{fig:normal_energy}.  {\it Panel
(b)} The redshift dependence of the partial fluence at $t=1.1e4$
seconds. The limiting redshifts are $z=0.25$ for \agile, and $z=0.8$
for \glast, respectively. {\it Panel (c)} The redshift dependence of
the partial fluence at $t=2.0 \times 10^4$ seconds.  The limiting
redshift points are $z=0.15$ for \agile, and $z=0.72$ for \glast,
respectively.  {\it Panel (d)} The redshift dependence of the partial
fluence at $t=1.4 \times 10^5$ seconds.  The limiting redshift for
\agile\, is below $z\sim 0.1$, and the limiting redshift for \glast\,
is around $z=0.5$, for this model.}
\label{fig:agile_glast}
\end{figure}

In Fig \ref{fig:agile_glast} we show the corresponding results for
\agile. This is a smaller-scale mission than \glast, launched in April
23, 2007, and it is interesting to compare its detectability limits
with those of \glast.  \agile\, has a different energy range (30 MeV
to 50 GeV) and has a relatively narrower energy band than the \glast\,
LAT (20 MeV to 300 GeV), as well as a lower effective area.  Thus the
observed fluxes and partial fluences are expected to be lower for
\agile\, in contrast to \glast. This is seen in Panel (a) of Fig
\ref{fig:agile_glast}. The dashed line is for \agile, and the solid
line is for \glast, showing that it is hard for \agile\, to detect a
burst with the typical parameters at $z=0.32$, while \glast\, could
detect it until around 2 days. Panels (b), (c) and (d) of Fig
\ref{fig:agile_glast} show the detectability with \agile\, and with
\glast\, at different time epochs.  At $t=1.1\times 10^3$ seconds,
\agile\, can detect bursts up to $z\simeq 0.25$, and \glast\, can
detect bursts to $z\simeq 0.8$.  At the time $t=2.0\times 10^4$
seconds, the limiting redshifts are $0.15$ for \agile, and $0.8$ for
\glast (same as Fig \ref{fig:normal_energy}), respectively.  At $t=1.4
\times 10^5$ seconds, the limiting redshift for \agile\, is apparently
well below $z=0.1$, while the limiting redshift for \glast\, can still
reach up to $0.5$.  We see that the limiting redshift for \agile\,
drops relatively quickly with increasing observation times; the
short-time sensitivity for \agile\, lasts around $10^3$ seconds, and
after that the sensitivity drops quickly. For \glast, the short-time
sensitivity lasts a longer time, $\sim10^4$ seconds, and since the
afterglow GeV flux doesn't change much for up to one day after the
trigger, we don't expect the sensitivity drop to have much of an
effect on the limiting redshift for \glast.

\begin{figure}
\plotone{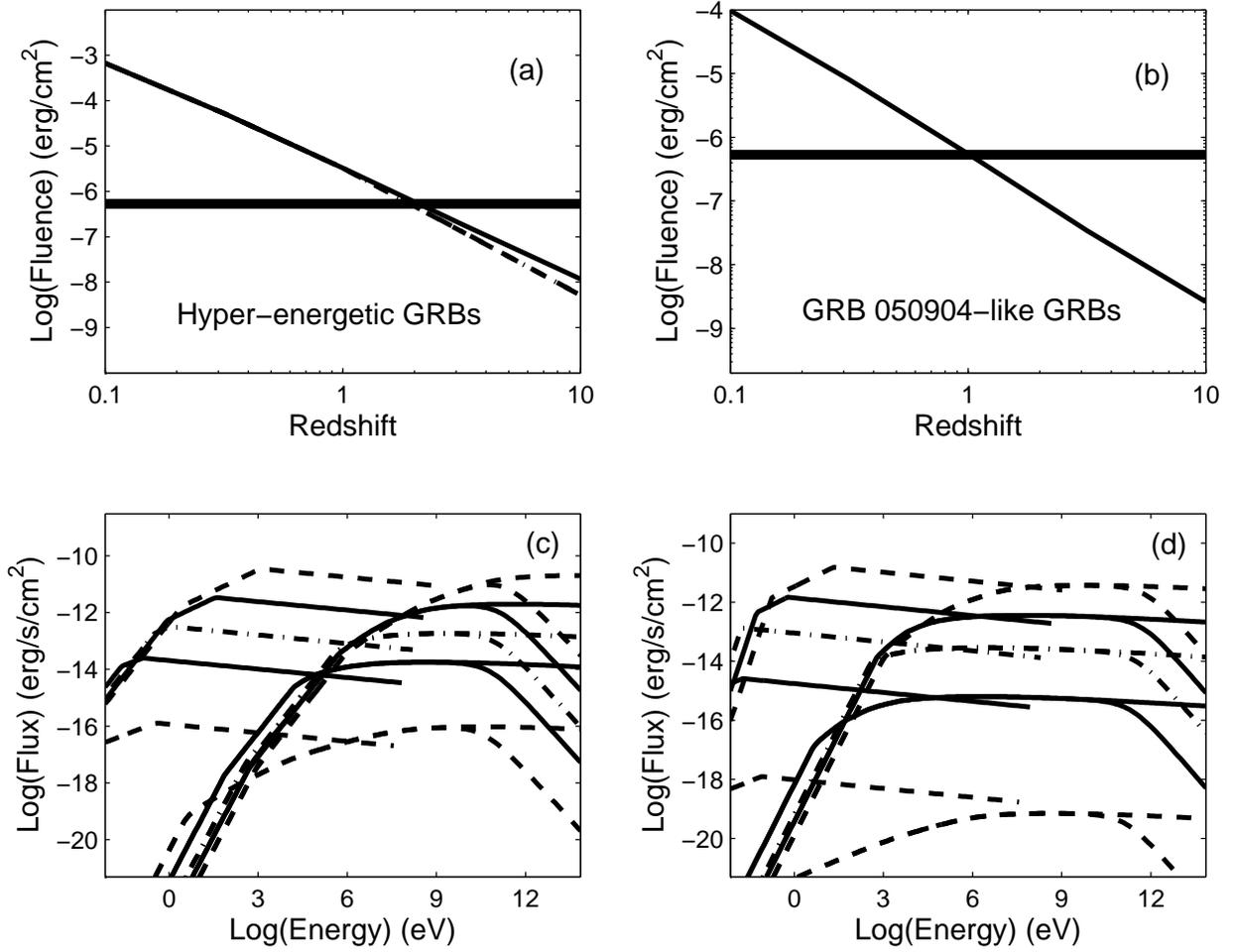}
\caption{Standard models of type (A), with different values of the
parameters.  {\it Panel (a)} The redshift dependence of the GeV partial fluence
for hyper-energetic GRB 990123-like objects, at $t=2.0\times 10^4$
seconds.  The intersection point gives the limiting redshift $z\simeq
2$ below which GLAST can detect such afterglows in the GeV band. The
model parameters are $E_{52,iso}=100$,~$\epsilon_e=0.2$ and the other
parameters are the same as in Fig (\ref{fig:zhang_detection}).  {\it
Panel (b)}: The redshift dependence of the partial fluence for model (B) of GRB
050904 using the best fitting for this bursts parameters
\citep{gfm+06}, $p=2.194, \epsilon_e=0.026, \epsilon_B=0.0058, n=109,
\theta=0.111, E_{52,iso}=184.8$, at the time $t=2.0\times 10^4$
seconds.  The limiting redshift for such GRB 050904-like objects is
$z\simeq 1$.  {\it Panel (c)} The synchrotron and IC spectrum for
hyper-energetic GRB 990123-like objects. {\it Panel (d)} The
synchrotron and IC spectrum for GRB 050904-like GRBs. }
\label{fig:high_energy}
\end{figure}

In Figure \ref{fig:high_energy} we probe the sensitivity of the
detectabilty on the total kinetic energy of the burst, taking as an
example the results for a value of $E_{52,iso}=100$. This is in the
range of values derived for objects such as GRB 990123 and GRB 050904,
which may be called hyper-energetic GRB. For the ``standard'' model
case (A) with this energy (see upper panels of Fig
\ref{fig:high_energy}) we see that the limiting redshift for a
\glast\, detection has increased from $z\simeq 0.8$ to a value
$z\simeq 2.0$, assuming that the other parameters remain the same as
in Figure \ref{fig:normal_energy}. Thus, hyper-energetic bursts such
as GRB 990123, at the observed redshift $z=1.6$, should be detected by
GLAST in the GeV band, if they have the above conventional
parameters. The other hyper-energetic object, GRB 050904, has a
similar kinetic energy as GRB 990123, but it was at the much higher
redshift $z=6.29$, which appears out of range for GLAST. GRB 050904
had the most complete set of observational data so far, covering from
the BAT band, through X-ray, to the optical/NIR and to the radio
band. Thus a lot of effort has been invested in obtaining the best
fitting parameters for this burst \citep{fckn+06, gfm+06}. Taking the
best fitting parameters from model (B) of \citet{gfm+06}, our results
here indicate that such GRB 050904-like bursts could be detected by
\glast\, up to $z \le 1.0$. There are two reasons for this relatively
modest limiting redshift detectability by GLAST in this case: (1) The
electron equipartition parameter derived is small, $\epsilon_e=0.026$,
which means only a fraction of the kinetic energy is radiated; (2) The
Compton parameter parameter in the fast cooling case is relatively
small, $Y\simeq 1.7$, which means that the energy lost via IC
scattering is comparable to the energy lost via synchrotron radiation.

\begin{figure}
\plotone{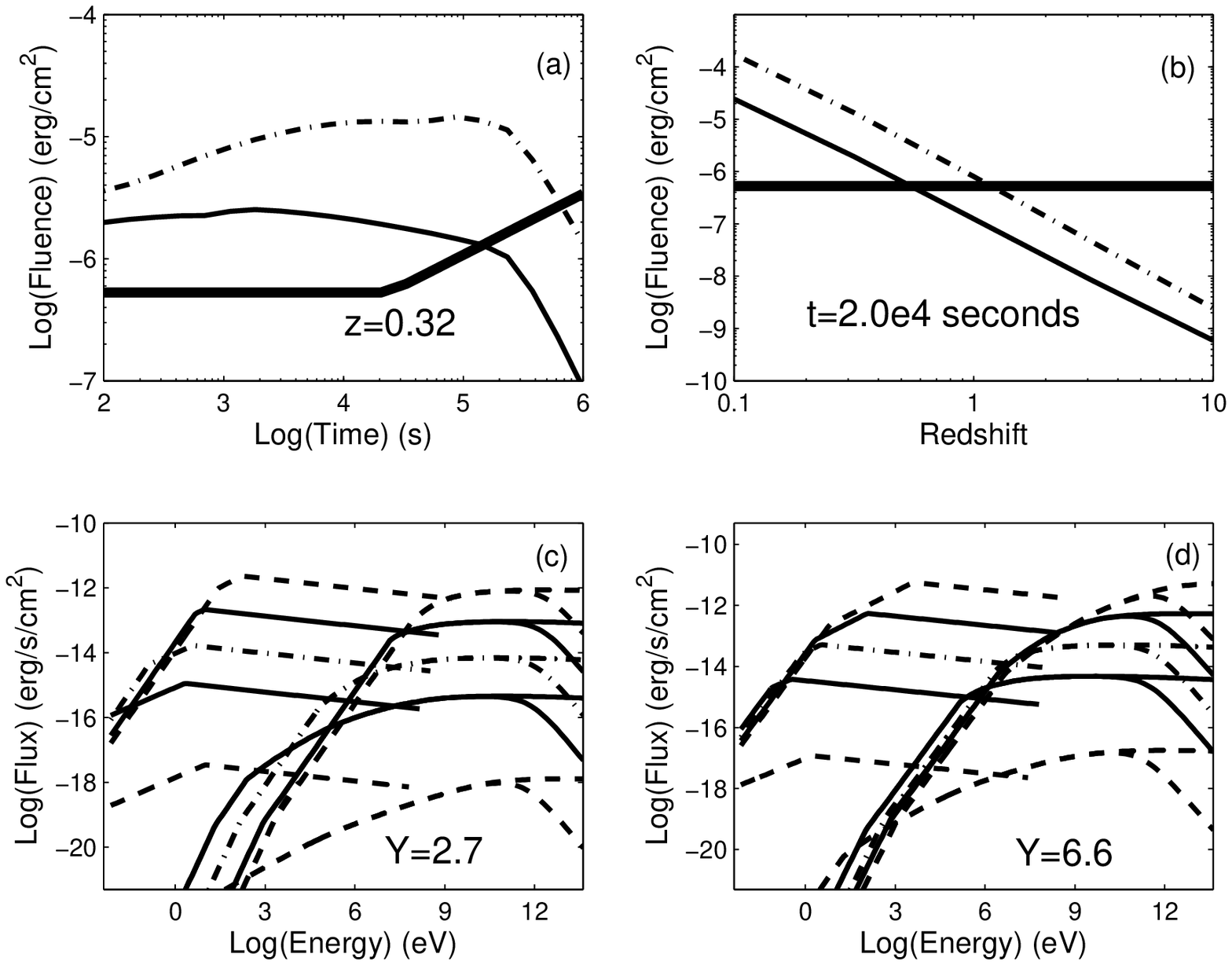}
\caption{Detectability limits for \glast\, for standard models such as
(A) but with different Compton Y parameters, $Y=2.7$ and $Y=6.6$. {\it
Panel (a)} The partial fluence curves for $Y=6.6$ (thin dashed dotted
line) and $Y=2.7$ (thin solid line), compared to the GLAST sensitivity
(thick broken solid line). {\it Panel (b)} The limiting redshift for
an observation time $t=2\times 10^4$ s and the two Y parameters,
giving limiting redshifts $z=0.55$, and $z=1.2$, respectively. {\it
Panel (c)} The synchrotron and IC spectra for the Compton parameter
$Y=2.7$ case.  {\it Panel (d)} The synchrotron and IC spectra for the
Compton parameter $Y=6.6$ case.} \label{fig:different_y}
\end{figure}

In Figure \ref{fig:different_y} we illustrate the sensitivity of the
GeV detectability on the value of the Compton Y parameter, showing the
fluxes for two values, $Y=2.7$ (thin solid line) and $Y=6.6$ (thin
dashed dotted line) in the fast cooling case.  The parameters are same as the
ones of the standard case (A) in Fig. \ref{fig:normal_energy}, except
for the electron equipartition parameter $\epsilon_e$, which is
$\epsilon_e=0.1$ for the $Y=2.7$ case and $\epsilon_e=0.5$ for the
$Y=6.6$ case. One expects a higher flux over the \glast\, energy band
for the $Y=6.6$ case because a larger Compton parameter means that
more energy goes into the GeV band via the IC process, which can be
seen from the spectrum on the lower panels.  The limiting redshifts
for the observation time $t=2.0\times 10^4$ seconds are $z=0.55$ and
$z=1.2$, respectively, for $Y=2.7$ and $Y=6.6$.

\section{Discussion}
\label{sec:disc}

We have calculated the time-dependent GeV synchrotron and inverse
Compton spectra for three generic types of GRB afterglows which, at
lower photon energies, have been used to interpret observations from
the Swift satellite and other ground-based facilities. These models
include a standard, constant parameter afterglow model (A), a model
with late energy injection (B) and a model with varying parameters
(C). The spectra and the GeV partial fluence curves in the GeV range
were used to estimate the detectability with \glast\, and \agile\, of
bursts in these model categories, for various sets of parameters, at
various epochs after the trigger and for various observation
durations.  These model spectra improve on previous calculations in
several respects.  In particular, in the past mainly constant
parameter spectral models such as (A) were computed; here we have
extended these calculations over a broader range of input parameters,
based on more recent information and statistics. Models (B) and (C)
have not been previously investigated in the GeV range, and are
motivated by recent Swift results.  The GeV spectra discussed here
were computed numerically, using the formalism described in
\citet{gfm+06}. These are compared with previous analytical
synchrotron-IC spectra of type (A) in the appendix.

The detectability depends most obviously on the total burst energy
$E_{52,iso}$ and on the observation time $t$ and integration time $\Delta t$, 
in addition to the other parameters such as $\epsilon_e$, $\epsilon_B$
etc. E.g., for bursts of the standard constant parameters (A) with the
nominal values of the parameters $E_{52,iso}=1$, $\epsilon_e=0.5$,
$\epsilon_B=0.01$, the limiting detection redshift with \glast\, for
times $t=1.1\times 10^3$ s and $2\times 10^4$ s and integration time
$\Delta t=0.5t$ are roughly the same,
$z\simeq 0.4$ (Fig. \ref{fig:zhang_detection}, panels b and c).  For
models with energy injection (B) or varying parameters (C), where the
final values of $E_{iso}$ or $\epsilon_e$ reach the same value as for
model (A) at a later time, the detection threshold is somewhat lower
for the shorter observation time, as seen in the same figure, panel (c),
since they start out weaker and build up to be comparable to (A) at
later times. However, for the longer of the two observation times above,
panel d, the limiting redshifts are the same.  For the more standard
values $E_{52,iso}=10$, $\epsilon_e=0.2$ (Fig.
\ref{fig:normal_energy}, panel b) the limiting redshift for the
constant parameter model (A) at $t=2\times 10^4$ s goes up to $z\simeq
0.8$.

For \agile\, the limiting redshifts are lower than for \glast\, due to
its lower effective area.  For a standard burst model (A) with average
parameter values $E_{52,iso}=10$, $\epsilon_e=0.2$, the limiting
redshift is $z\simeq 0.25$ at an earlier observation time $t=1.1\times 10^3$ s
(Fig. \ref{fig:agile_glast}, panel b), and $z\simeq 0.15$ at
$t=2\times 10^4$ s. With \agile\, bursts can only be detected at
relatively early times, since the short-time sensitivity for \agile\,
(where the sensitivity curve is flat in time) only lasts, e.g. around
$\sim 10^3$ seconds, versus $\sim 10^4$ s for \glast\, for a burst at
$z\sim 0.3$ (Fig.  \ref{fig:agile_glast}, panel a).

A discrimination between models (A), (B) and (C) based on GeV
measurements of the spectral evolution in time is possible in
principle, as seen e.g. by comparing Fig. 1 panel (c), and Fig. 2
panels (c) and (d).  However, it will require good energy and time
coverage, and extensive simulations over a wide range of parameter
space, since changes in $E_{iso}$, $Y$
(e.g. Figs. \ref{fig:high_energy} and \ref{fig:different_y}) and the
other afterglow parameters needs to be carefully disentangled. 

The limiting redshift naturally increases for larger values of
$E_{iso}$. E.g., for \glast\, at an observation time $t=2\times
10^4$ s and a standard model (A) with $E_{52,iso}=10^2$,
$\epsilon_e=0.2$ it is $z\simeq 2$, while for a for a burst with the
parameters of GRB 050904 (high $E_{iso}$ but low $\epsilon_e$) it is
$z\simeq 1$. The limiting redshift also increases with the Compton $Y$
parameter, as illustrated in Fig. \ref{fig:different_y}, which
reflects the fact that $Y$ provides a measure of how much energy gets
scattered into the GeV range. 

Besides the photon-photon absorption inside the source considered in
our calculation, another interesting absorption process is the
photon-photon absorption by the external cosmic infrared background
radiation (external absorption), which produces electron/positron
pairs, and the resulting pairs can IC scatter the cosmic microwave
background (CMB) photons, yielding a delayed MeV-GeV emission
\citep{dl+02}. It can be shown that the external absorption doesn't
affect the limiting redshift much. \citet{ss+98} show that, at $z=1$,
the absorption optical depth is $\tau\simeq 1$ for photons with
an energy of 50 GeV, and $\tau \simeq 10$ for photons of 300 GeV. 
If we define the cutoff energy as the photon energy corresponding to 
an optical depth $\tau=1$, the cutoff energy should shift to higher 
energies as one considers lower redshift, based on the numerical 
results in \citet{ss+98}. Assuming that (1) the photons above 50 GeV 
is totally absorbed and there is no absorption below 50 GeV; 
(2) the cutoff energy is roughly the same within the redshift range 
considered; and (3) the flux contribution from the delayed emission is 
ignored, we can conclude that in this case the flux observed by \glast\,
will be comparable to that observed by \agile, because the effective 
observing band will have become similar for both instruments (due to
external absorption effectively cutting off the \glast\, higher energy 
contribution). The limiting redshift for \agile, as shown in Fig 
\ref{fig:agile_glast}, will be approximately the same for the case with 
external absorption as it is without, because absorption is important
only above the \agile\, band.

The detectability estimates discussed here illustrate the sensitivity
to different types of model assumptions in current synchrotron-inverse
Compton models, when one takes into account newer information gleaned
from \swift. Calculations using simplified generic models show that
around tens of \swift -detected GRB per year will fall in the LAT
field of view \citep{omo+06} during their prompt emission
phase. Considering the \swift-detected burst redshift distribution to
imply a fraction of around $20\%$ below $z=1$ \citep{jlfp+06,ld+06},
we may roughly expect $\sim 5$ \swift-burst prompt detections per year
by \glast. However, since the GeV afterglows can last up to a day
(e.g. Figs. 1 and 2, panel a), \glast\, may actually be able to observe
more than this number of bursts in the afterglow, as opposed to the
prompt phase.  Detections with \glast\, should test many of the
assumptions that go into these models, and will provide important new
information on the energetics, dynamics and parameters of GRB
afterglows.

\acknowledgements

We are grateful to J. McEnery for informative correspondence, and
Z.G. Dai, C. Dermer, D. Fox, S. Kobayashi, L. Stella, X.Y. Wang, and
B. Zhang for useful comments. This research has been supported in part
through NASA NAG5-13286 and NSF AST 0307376.  L.~J. Gou also thanks
support from the Sigma-Xi Fellowship.

\section{Appendix}
\label{sec:appendix}

We discuss here two analytical approximations to the synchrotron-IC
spectrum, and compare them to the numerically calculated values. The
two key elements in the simple analytical approximations to
synchrotron-IC spectra in the literature are (i) an IC-to-synchrotron
peak flux ratio, \beq F \equiv f^{IC}_{max}/f^{syn}_{max}
\label{eq:fdef} \enq 
expressed, e.g. in erg cm$^{-2}$ s$^{-1}$ Hz$^{-1}$ and evaluated at
the frequencies where the synchrotron and the IC flux attain their
peak value, and (ii) a ``Compton parameter'' Y, usually taken to be
\beq Y=(-1+\sqrt{4\epsilon_e/\epsilon_B+1})/2 ~. \label{eq:y-param}
\enq In the GRB literature, the flux ratio (\ref{eq:fdef}) of the
analytical approximations has appeared under several forms, two of
which are the most relevant for us here. One of these is \beq F1\equiv
f^{IC}_{max}/f^{syn}_{max}=(14/45)\sigma_T R n \simeq \tau
\label{eq:fsariesin} \enq 
\citep*[see][A9]{se+01}, where $\sigma_T$ is the Thomson scattering
cross section, $R$ is the shock radius, $n$ is the external
circumburst density in units of $\rm cm^{-3}$, and $\tau= \sigma_T R
n/3$ is the Thomson optical depth of the radiation region
\footnote{The optical depth here differs from the usual definition by
a factor 1/3, but for consistency with the usage in the literature
\citep{pk+00,kzmb+07} we keep the factor 1/3 here.}.  Another form of
this ratio which has been used is \beq F2 \equiv
f^{IC}_{max}/f^{syn}_{max}=[4(p-1)/(p-2)] \tau ~,
\label{eq:fkoba} 
\enq \citep{kzmb+07}. It is apparent that F1 is smaller than F2 by a
factor $[4(p-1)/(p-2)]$, which can give substantial differences in
analytical estimates of the IC spectral flux at its peak. It is
worthwhile therefore to clarify the reason for the discrepancy.

The F1 form (equation [\ref{eq:fsariesin}]) of the peak flux ratio is
derived from an integral over the electron energy distribution and a
power-law seed synchrotron spectrum (See Eqn. (7.28),
\citeauthor{rl+79}~\citeyear{rl+79}; Also Eqn. (A1),
\citeauthor{se+01}~\citeyear{se+01}).

The F2 form (equation [\ref{eq:fkoba}]) of the peak flux ratio, on the other
hand, is obtained by solving for $f^{IC}_{max}/f^{syn}_{max}$ from an equation
(\ref{eq:lum-ratio}) which relates the Compton Y parameter to the ratio of
the luminosities produced by the first order IC and the synchrotron mechanisms
\citep{kzmb+07},
\beq
Y=L_{IC,1st}/L_{syn} \sim \nu^{IC}_{peak} f_{\nu}^{IC}(\nu^{IC}_{peak}) / \nu_{peak}
f_{\nu} (\nu_{peak})=2\kappa \tau \gamma_m \gamma_c
\label{eq:lum-ratio}
\enq where $\nu^{IC}_{peak}$ and $\nu_{peak}$ are the peak frequency
of the IC and synchrotron spectra, respectively, and
$f_{\nu}^{IC}(\nu^{IC}_{peak})$ and $f_{\nu} (\nu_{peak})$ are the
peak fluxes corresponding to the IC and synchrotron peak
frequencies. The corresponding analytical approximation to the IC
spectrum is a simpler one than in the previous F1 case, in that it is
a broken power law (without the logarithmic corrections). In the
low-frequency part of the IC spectrum, the broken power law is a good
approximation to a numerically calculated IC spectrum. However, in the
high-frequency part, the broken power-law analytical approximation
under-estimates the numerical IC spectrum, which is larger (and has a
flatter spectral index) than the broken power-law prediction (this
under-estimation is avoided in the F1 case by including the
logarithmic correction). In the F2 pure broken power law case,
therefore, in order to keep the frequency integrated total luminosity
radiated by the IC mechanism equal to the numerically computed total
IC luminosity, and to preserve the desirable simple broken power law
shape, an artificially boosted peak flux ratio is adopted, which leads
to the same IC-to-synchrotron luminosity ratio.  Thus, while the total
energetics are the same for both analytical approximations, the IC
flux expected over the \glast\, (and \agile) energy range
differ. Whereas $F2$ is simpler for quick estimates since it involves
pure power laws and correctly describes the global energetics, $F1$
with the logarithmic corrections to the power laws is preferable for
more accurate GeV spectral flux estimates.

\end{document}